\documentclass{eptcs}
\usepackage{underscore}           

\usepackage[capitalize,noabbrev]{cleveref}
\usepackage{graphicx}
\usepackage{lineno}
\usepackage{cite}
\usepackage{longtable}

\title{Teaching Functional Programmers Logic and Metatheory}

\author{Frederik Krogsdal Jacobsen \qquad Jørgen Villadsen
\institute{DTU Compute - Department of Applied Mathematics and Computer Science - Technical University of Denmark}
}

\def\authorrunning{F.\,K. Jacobsen and J. Villadsen}
\def\titlerunning{Teaching Functional Programmers Logic and Metatheory}

\usepackage{isabelle, isabellesym}
\isabellestyle{it} 

\newcommand{\DefineSnippet}[2]{\expandafter\newcommand\csname snippet--#1\endcsname{#2}}
\DefineSnippet{theory:Scratch-0}{%
\isadelimtheory
\endisadelimtheory
\isatagtheory
\isacommand{theory}\isamarkupfalse%
\ Scratch\ \isakeyword{imports}\ Main\ \isakeyword{begin}%
\endisatagtheory
{\isafoldtheory}%
\isadelimtheory
\endisadelimtheory
}
\DefineSnippet{datatype:form-0}{%
\isacommand{datatype}\isamarkupfalse%
\ {\isacharprime}{\kern0pt}a\ form\ \\\hspace*{5mm} {\isacharequal}{\kern0pt}\ Pro\ {\isacharprime}{\kern0pt}a\ {\isacharparenleft}{\kern0pt}{\isacartoucheopen}{\isasymcdot}{\isacartoucheclose}{\isacharparenright}{\kern0pt}\ {\isacharbar}{\kern0pt}\ Falsity\ {\isacharparenleft}{\kern0pt}{\isacartoucheopen}{\isasymbottom}{\isacartoucheclose}{\isacharparenright}{\kern0pt}\ {\isacharbar}{\kern0pt}\ Imp\ {\isacartoucheopen}{\isacharprime}{\kern0pt}a\ form{\isacartoucheclose}\ {\isacartoucheopen}{\isacharprime}{\kern0pt}a\ form{\isacartoucheclose}\ {\isacharparenleft}{\kern0pt}\isakeyword{infixr}\ {\isacartoucheopen}{\isasymrightarrow}{\isacartoucheclose}\ {\isadigit{0}}{\isacharparenright}{\kern0pt}%
}
\DefineSnippet{primrec:semantics-0}{%
\isacommand{primrec}\isamarkupfalse%
\ semantics\ \isakeyword{where}\isanewline
}
\DefineSnippet{primrec:semantics-1}{%
\ \ {\isacartoucheopen}semantics\ i\ {\isacharparenleft}{\kern0pt}{\isasymcdot}n{\isacharparenright}{\kern0pt}\ {\isacharequal}{\kern0pt}\ i\ n{\isacartoucheclose}\ {\isacharbar}{\kern0pt}\isanewline
}
\DefineSnippet{primrec:semantics-2}{%
\ \ {\isacartoucheopen}semantics\ {\isacharunderscore}{\kern0pt}\ {\isasymbottom}\ {\isacharequal}{\kern0pt}\ False{\isacartoucheclose}\ {\isacharbar}{\kern0pt}\isanewline
}
\DefineSnippet{primrec:semantics-3}{%
\ \ {\isacartoucheopen}semantics\ i\ {\isacharparenleft}{\kern0pt}p\ {\isasymrightarrow}\ q{\isacharparenright}{\kern0pt}\ {\isacharequal}{\kern0pt}\ {\isacharparenleft}{\kern0pt}semantics\ i\ p\ {\isasymlongrightarrow}\ semantics\ i\ q{\isacharparenright}{\kern0pt}{\isacartoucheclose}%
}
\DefineSnippet{abbreviation:sc-0}{%
\isacommand{abbreviation}\isamarkupfalse%
\ {\isacartoucheopen}sc\ X\ Y\ i\ {\isasymequiv}\ {\isacharparenleft}{\kern0pt}{\isasymforall}p\ {\isasymin}\ set\ X{\isachardot}{\kern0pt}\ semantics\ i\ p{\isacharparenright}{\kern0pt}\ {\isasymlongrightarrow}\ {\isacharparenleft}{\kern0pt}{\isasymexists}q\ {\isasymin}\ set\ Y{\isachardot}{\kern0pt}\ semantics\ i\ q{\isacharparenright}{\kern0pt}{\isacartoucheclose}%
}
\DefineSnippet{primrec:member-0}{%
\isacommand{primrec}\isamarkupfalse%
\ member\ \isakeyword{where}\isanewline
}
\DefineSnippet{primrec:member-1}{%
\ \ {\isacartoucheopen}member\ {\isacharunderscore}{\kern0pt}\ {\isacharbrackleft}{\kern0pt}{\isacharbrackright}{\kern0pt}\ {\isacharequal}{\kern0pt}\ False{\isacartoucheclose}\ {\isacharbar}{\kern0pt}\isanewline
}
\DefineSnippet{primrec:member-2}{%
\ \ {\isacartoucheopen}member\ m\ {\isacharparenleft}{\kern0pt}n\ {\isacharhash}{\kern0pt}\ A{\isacharparenright}{\kern0pt}\ {\isacharequal}{\kern0pt}\ {\isacharparenleft}{\kern0pt}m\ {\isacharequal}{\kern0pt}\ n\ {\isasymor}\ member\ m\ A{\isacharparenright}{\kern0pt}{\isacartoucheclose}%
}
\DefineSnippet{lemma:member-iff-0}{%
\isacommand{lemma}\isamarkupfalse%
\ member{\isacharunderscore}{\kern0pt}iff\ {\isacharbrackleft}{\kern0pt}iff{\isacharbrackright}{\kern0pt}{\isacharcolon}{\kern0pt}\ {\isacartoucheopen}member\ m\ A\ {\isasymlongleftrightarrow}\ m\ {\isasymin}\ set\ A{\isacartoucheclose}\isanewline
}
\DefineSnippet{lemma:member-iff-1}{%
\isadelimproof
\ \ %
\endisadelimproof
\isatagproof
\isacommand{by}\isamarkupfalse%
\ {\isacharparenleft}{\kern0pt}induct\ A{\isacharparenright}{\kern0pt}\ simp{\isacharunderscore}{\kern0pt}all%
\endisatagproof
{\isafoldproof}%
\isadelimproof
\endisadelimproof
}
\DefineSnippet{primrec:common-0}{%
\isacommand{primrec}\isamarkupfalse%
\ common\ \isakeyword{where}\isanewline
}
\DefineSnippet{primrec:common-1}{%
\ \ {\isacartoucheopen}common\ {\isacharunderscore}{\kern0pt}\ {\isacharbrackleft}{\kern0pt}{\isacharbrackright}{\kern0pt}\ {\isacharequal}{\kern0pt}\ False{\isacartoucheclose}\ {\isacharbar}{\kern0pt}\isanewline
}
\DefineSnippet{primrec:common-2}{%
\ \ {\isacartoucheopen}common\ A\ {\isacharparenleft}{\kern0pt}m\ {\isacharhash}{\kern0pt}\ B{\isacharparenright}{\kern0pt}\ {\isacharequal}{\kern0pt}\ {\isacharparenleft}{\kern0pt}member\ m\ A\ {\isasymor}\ common\ A\ B{\isacharparenright}{\kern0pt}{\isacartoucheclose}%
}
\DefineSnippet{lemma:common-iff-0}{%
\isacommand{lemma}\isamarkupfalse%
\ common{\isacharunderscore}{\kern0pt}iff\ {\isacharbrackleft}{\kern0pt}iff{\isacharbrackright}{\kern0pt}{\isacharcolon}{\kern0pt}\ {\isacartoucheopen}common\ A\ B\ {\isasymlongleftrightarrow}\ set\ A\ {\isasyminter}\ set\ B\ {\isasymnoteq}\ {\isacharbraceleft}{\kern0pt}{\isacharbraceright}{\kern0pt}{\isacartoucheclose}\isanewline
}
\DefineSnippet{lemma:common-iff-1}{%
\isadelimproof
\ \ %
\endisadelimproof
\isatagproof
\isacommand{by}\isamarkupfalse%
\ {\isacharparenleft}{\kern0pt}induct\ B{\isacharparenright}{\kern0pt}\ simp{\isacharunderscore}{\kern0pt}all%
\endisatagproof
{\isafoldproof}%
\isadelimproof
\endisadelimproof
}
\DefineSnippet{function:mp-0}{%
\isacommand{function}\isamarkupfalse%
\ mp\ \isakeyword{where}\isanewline
}
\DefineSnippet{function:mp-1}{%
\ \ {\isacartoucheopen}mp\ A\ B\ {\isacharparenleft}{\kern0pt}{\isasymcdot}n\ {\isacharhash}{\kern0pt}\ C{\isacharparenright}{\kern0pt}\ {\isacharbrackleft}{\kern0pt}{\isacharbrackright}{\kern0pt}\ {\isacharequal}{\kern0pt}\ mp\ {\isacharparenleft}{\kern0pt}n\ {\isacharhash}{\kern0pt}\ A{\isacharparenright}{\kern0pt}\ B\ C\ {\isacharbrackleft}{\kern0pt}{\isacharbrackright}{\kern0pt}{\isacartoucheclose}\ {\isacharbar}{\kern0pt}\isanewline
}
\DefineSnippet{function:mp-2}{%
\ \ {\isacartoucheopen}mp\ A\ B\ C\ {\isacharparenleft}{\kern0pt}{\isasymcdot}n\ {\isacharhash}{\kern0pt}\ D{\isacharparenright}{\kern0pt}\ {\isacharequal}{\kern0pt}\ mp\ A\ {\isacharparenleft}{\kern0pt}n\ {\isacharhash}{\kern0pt}\ B{\isacharparenright}{\kern0pt}\ C\ D{\isacartoucheclose}\ {\isacharbar}{\kern0pt}\isanewline
}
\DefineSnippet{function:mp-3}{%
\ \ {\isacartoucheopen}mp\ {\isacharunderscore}{\kern0pt}\ {\isacharunderscore}{\kern0pt}\ {\isacharparenleft}{\kern0pt}{\isasymbottom}\ {\isacharhash}{\kern0pt}\ {\isacharunderscore}{\kern0pt}{\isacharparenright}{\kern0pt}\ {\isacharbrackleft}{\kern0pt}{\isacharbrackright}{\kern0pt}\ {\isacharequal}{\kern0pt}\ True{\isacartoucheclose}\ {\isacharbar}{\kern0pt}\isanewline
}
\DefineSnippet{function:mp-4}{%
\ \ {\isacartoucheopen}mp\ A\ B\ C\ {\isacharparenleft}{\kern0pt}{\isasymbottom}\ {\isacharhash}{\kern0pt}\ D{\isacharparenright}{\kern0pt}\ {\isacharequal}{\kern0pt}\ mp\ A\ B\ C\ D{\isacartoucheclose}\ {\isacharbar}{\kern0pt}\isanewline
}
\DefineSnippet{function:mp-5}{%
\ \ {\isacartoucheopen}mp\ A\ B\ {\isacharparenleft}{\kern0pt}{\isacharparenleft}{\kern0pt}p\ {\isasymrightarrow}\ q{\isacharparenright}{\kern0pt}\ {\isacharhash}{\kern0pt}\ C{\isacharparenright}{\kern0pt}\ {\isacharbrackleft}{\kern0pt}{\isacharbrackright}{\kern0pt}\ {\isacharequal}{\kern0pt}\ {\isacharparenleft}{\kern0pt}mp\ A\ B\ C\ {\isacharbrackleft}{\kern0pt}p{\isacharbrackright}{\kern0pt}\ {\isasymand}\ mp\ A\ B\ {\isacharparenleft}{\kern0pt}q\ {\isacharhash}{\kern0pt}\ C{\isacharparenright}{\kern0pt}\ {\isacharbrackleft}{\kern0pt}{\isacharbrackright}{\kern0pt}{\isacharparenright}{\kern0pt}{\isacartoucheclose}\ {\isacharbar}{\kern0pt}\isanewline
}
\DefineSnippet{function:mp-6}{%
\ \ {\isacartoucheopen}mp\ A\ B\ C\ {\isacharparenleft}{\kern0pt}{\isacharparenleft}{\kern0pt}p\ {\isasymrightarrow}\ q{\isacharparenright}{\kern0pt}\ {\isacharhash}{\kern0pt}\ D{\isacharparenright}{\kern0pt}\ {\isacharequal}{\kern0pt}\ mp\ A\ B\ {\isacharparenleft}{\kern0pt}p\ {\isacharhash}{\kern0pt}\ C{\isacharparenright}{\kern0pt}\ {\isacharparenleft}{\kern0pt}q\ {\isacharhash}{\kern0pt}\ D{\isacharparenright}{\kern0pt}{\isacartoucheclose}\ {\isacharbar}{\kern0pt}\isanewline
}
\DefineSnippet{function:mp-7}{%
\ \ {\isacartoucheopen}mp\ A\ B\ {\isacharbrackleft}{\kern0pt}{\isacharbrackright}{\kern0pt}\ {\isacharbrackleft}{\kern0pt}{\isacharbrackright}{\kern0pt}\ {\isacharequal}{\kern0pt}\ common\ A\ B{\isacartoucheclose}\isanewline
}
\DefineSnippet{function:mp-8}{%
\isadelimproof
\ \ %
\endisadelimproof
\isatagproof
\isacommand{by}\isamarkupfalse%
\ pat{\isacharunderscore}{\kern0pt}completeness\ simp{\isacharunderscore}{\kern0pt}all%
\endisatagproof
{\isafoldproof}%
\isadelimproof
\endisadelimproof
}
\DefineSnippet{termination:mp-0}{%
\isacommand{termination}\isamarkupfalse%
\isanewline
}
\DefineSnippet{termination:mp-1}{%
\isadelimproof
\ \ %
\endisadelimproof
\isatagproof
\isacommand{by}\isamarkupfalse%
\ {\isacharparenleft}{\kern0pt}relation\ {\isacartoucheopen}measure\ {\isacharparenleft}{\kern0pt}{\isasymlambda}{\isacharparenleft}{\kern0pt}{\isacharunderscore}{\kern0pt}{\isacharcomma}{\kern0pt}\ {\isacharunderscore}{\kern0pt}{\isacharcomma}{\kern0pt}\ C{\isacharcomma}{\kern0pt}\ D{\isacharparenright}{\kern0pt}{\isachardot}{\kern0pt}\ {\isasymSum}p\ {\isasymleftarrow}\ C\ {\isacharat}{\kern0pt}\ D{\isachardot}{\kern0pt}\ size\ p{\isacharparenright}{\kern0pt}{\isacartoucheclose}{\isacharparenright}{\kern0pt}\ simp{\isacharunderscore}{\kern0pt}all%
\endisatagproof
{\isafoldproof}%
\isadelimproof
\endisadelimproof
}
\DefineSnippet{theorem:main-0}{%
\isacommand{theorem}\isamarkupfalse%
\ main{\isacharcolon}{\kern0pt}\ {\isacartoucheopen}{\isacharparenleft}{\kern0pt}{\isasymforall}i{\isachardot}{\kern0pt}\ sc\ {\isacharparenleft}{\kern0pt}map\ {\isasymcdot}\ A\ {\isacharat}{\kern0pt}\ C{\isacharparenright}{\kern0pt}\ {\isacharparenleft}{\kern0pt}map\ {\isasymcdot}\ B\ {\isacharat}{\kern0pt}\ D{\isacharparenright}{\kern0pt}\ i{\isacharparenright}{\kern0pt}\ {\isasymlongleftrightarrow}\ mp\ A\ B\ C\ D{\isacartoucheclose}\isanewline
}
\DefineSnippet{theorem:main-1}{%
\isadelimproof
\ \ %
\endisadelimproof
\isatagproof
\isacommand{by}\isamarkupfalse%
\ {\isacharparenleft}{\kern0pt}induct\ rule{\isacharcolon}{\kern0pt}\ mp{\isachardot}{\kern0pt}induct{\isacharparenright}{\kern0pt}\ {\isacharparenleft}{\kern0pt}simp{\isacharunderscore}{\kern0pt}all{\isacharcomma}{\kern0pt}\ blast{\isacharcomma}{\kern0pt}\ meson{\isacharcomma}{\kern0pt}\ fast{\isacharparenright}{\kern0pt}%
\endisatagproof
{\isafoldproof}%
\isadelimproof
\endisadelimproof
}
\DefineSnippet{definition:prover-0}{%
\isacommand{definition}\isamarkupfalse%
\ {\isacartoucheopen}prover\ p\ {\isasymequiv}\ mp\ {\isacharbrackleft}{\kern0pt}{\isacharbrackright}{\kern0pt}\ {\isacharbrackleft}{\kern0pt}{\isacharbrackright}{\kern0pt}\ {\isacharbrackleft}{\kern0pt}{\isacharbrackright}{\kern0pt}\ {\isacharbrackleft}{\kern0pt}p{\isacharbrackright}{\kern0pt}{\isacartoucheclose}%
}
\DefineSnippet{corollary:13947c02cc06a4ca-0}{%
\isacommand{corollary}\isamarkupfalse%
\ {\isacartoucheopen}prover\ p\ {\isasymlongleftrightarrow}\ {\isacharparenleft}{\kern0pt}{\isasymforall}i{\isachardot}{\kern0pt}\ semantics\ i\ p{\isacharparenright}{\kern0pt}{\isacartoucheclose}\isanewline
}
\DefineSnippet{corollary:13947c02cc06a4ca-1}{%
\isadelimproof
\ \ %
\endisadelimproof
\isatagproof
\isacommand{unfolding}\isamarkupfalse%
\ prover{\isacharunderscore}{\kern0pt}def\ \isacommand{by}\isamarkupfalse%
\ {\isacharparenleft}{\kern0pt}simp\ flip{\isacharcolon}{\kern0pt}\ main{\isacharparenright}{\kern0pt}%
\endisatagproof
{\isafoldproof}%
\isadelimproof
\endisadelimproof
}
\DefineSnippet{abbreviation:neg-0}{%
\isacommand{abbreviation}\isamarkupfalse%
\ {\isacartoucheopen}neg\ a\ {\isasymequiv}\ {\isacharparenleft}{\kern0pt}a\ {\isasymrightarrow}\ {\isasymbottom}{\isacharparenright}{\kern0pt}{\isacartoucheclose}%
}
\DefineSnippet{abbreviation:con-0}{%
\isacommand{abbreviation}\isamarkupfalse%
\ {\isacartoucheopen}con\ a\ b\ {\isasymequiv}\ neg\ {\isacharparenleft}{\kern0pt}a\ {\isasymrightarrow}\ neg\ b{\isacharparenright}{\kern0pt}{\isacartoucheclose}%
}
\DefineSnippet{abbreviation:bii-0}{%
\isacommand{abbreviation}\isamarkupfalse%
\ {\isacartoucheopen}bii\ a\ b\ {\isasymequiv}\ con\ {\isacharparenleft}{\kern0pt}a\ {\isasymrightarrow}\ b{\isacharparenright}{\kern0pt}\ {\isacharparenleft}{\kern0pt}b\ {\isasymrightarrow}\ a{\isacharparenright}{\kern0pt}{\isacartoucheclose}%
}
\DefineSnippet{abbreviation:one-0}{%
\isacommand{abbreviation}\isamarkupfalse%
\ {\isacartoucheopen}one\ a\ b\ c\ {\isasymequiv}\isanewline
}
\DefineSnippet{abbreviation:one-1}{%
\ \ con\isanewline
}
\DefineSnippet{abbreviation:one-2}{%
\ \ \ \ {\isacharparenleft}{\kern0pt}bii\ a\ {\isacharparenleft}{\kern0pt}con\ {\isacharparenleft}{\kern0pt}neg\ b{\isacharparenright}{\kern0pt}\ {\isacharparenleft}{\kern0pt}neg\ c{\isacharparenright}{\kern0pt}{\isacharparenright}{\kern0pt}{\isacharparenright}{\kern0pt}\isanewline
}
\DefineSnippet{abbreviation:one-3}{%
\ \ \ \ {\isacharparenleft}{\kern0pt}con\isanewline
}
\DefineSnippet{abbreviation:one-4}{%
\ \ \ \ \ \ {\isacharparenleft}{\kern0pt}bii\ b\ {\isacharparenleft}{\kern0pt}con\ {\isacharparenleft}{\kern0pt}neg\ a{\isacharparenright}{\kern0pt}\ {\isacharparenleft}{\kern0pt}neg\ c{\isacharparenright}{\kern0pt}{\isacharparenright}{\kern0pt}{\isacharparenright}{\kern0pt}\isanewline
}
\DefineSnippet{abbreviation:one-5}{%
\ \ \ \ \ \ {\isacharparenleft}{\kern0pt}bii\ c\ {\isacharparenleft}{\kern0pt}con\ {\isacharparenleft}{\kern0pt}neg\ a{\isacharparenright}{\kern0pt}\ {\isacharparenleft}{\kern0pt}neg\ b{\isacharparenright}{\kern0pt}{\isacharparenright}{\kern0pt}{\isacharparenright}{\kern0pt}{\isacharparenright}{\kern0pt}{\isacartoucheclose}%
}
\DefineSnippet{datatype:people-0}{%
\isacommand{datatype}\isamarkupfalse%
\ people\ {\isacharequal}{\kern0pt}\ Ann\ {\isacharbar}{\kern0pt}\ Bob\ {\isacharbar}{\kern0pt}\ Cat%
}
\DefineSnippet{abbreviation:riddle-0}{%
\isacommand{abbreviation}\isamarkupfalse%
\ {\isacartoucheopen}riddle\ {\isasymequiv}\ {\isacharparenleft}{\kern0pt}\isanewline
}
\DefineSnippet{abbreviation:riddle-1}{%
\ \ con\isanewline
}
\DefineSnippet{abbreviation:riddle-2}{%
\ \ \ \ {\isacharparenleft}{\kern0pt}{\isasymcdot}Ann\ {\isasymrightarrow}\ bii\ {\isacharparenleft}{\kern0pt}{\isasymcdot}Bob{\isacharparenright}{\kern0pt}\ {\isacharparenleft}{\kern0pt}one\ {\isacharparenleft}{\kern0pt}{\isasymcdot}Ann{\isacharparenright}{\kern0pt}\ {\isacharparenleft}{\kern0pt}{\isasymcdot}Bob{\isacharparenright}{\kern0pt}\ {\isacharparenleft}{\kern0pt}{\isasymcdot}Cat{\isacharparenright}{\kern0pt}{\isacharparenright}{\kern0pt}{\isacharparenright}{\kern0pt}\isanewline
}
\DefineSnippet{abbreviation:riddle-3}{%
\ \ \ \ {\isacharparenleft}{\kern0pt}bii\ {\isacharparenleft}{\kern0pt}{\isasymcdot}Ann{\isacharparenright}{\kern0pt}\ {\isacharparenleft}{\kern0pt}neg\ {\isacharparenleft}{\kern0pt}{\isasymcdot}Cat{\isacharparenright}{\kern0pt}{\isacharparenright}{\kern0pt}{\isacharparenright}{\kern0pt}{\isacharparenright}{\kern0pt}{\isacartoucheclose}%
}
\DefineSnippet{proposition:cf9fa75d11c29c6d-0}{%
\isacommand{proposition}\isamarkupfalse%
\ {\isacartoucheopen}prover\ {\isacharparenleft}{\kern0pt}riddle\ {\isasymrightarrow}\ neg\ {\isacharparenleft}{\kern0pt}{\isasymcdot}Ann{\isacharparenright}{\kern0pt}{\isacharparenright}{\kern0pt}{\isacartoucheclose}%
\isadelimproof
\ %
\endisadelimproof
\isatagproof
\isacommand{by}\isamarkupfalse%
\ eval%
\endisatagproof
{\isafoldproof}%
\isadelimproof
\endisadelimproof
}
\DefineSnippet{proposition:2ecccc3cbfdfe1f5-0}{%
\isacommand{proposition}\isamarkupfalse%
\ {\isacartoucheopen}prover\ {\isacharparenleft}{\kern0pt}riddle\ {\isasymrightarrow}\ {\isasymcdot}Cat{\isacharparenright}{\kern0pt}{\isacartoucheclose}%
\isadelimproof
\ %
\endisadelimproof
\isatagproof
\isacommand{by}\isamarkupfalse%
\ eval%
\endisatagproof
{\isafoldproof}%
\isadelimproof
\endisadelimproof
}
\DefineSnippet{proposition:2f510588577c5e16-0}{%
\isacommand{proposition}\isamarkupfalse%
\ {\isacartoucheopen}{\isasymnot}prover\ {\isacharparenleft}{\kern0pt}riddle\ {\isasymrightarrow}\ neg\ {\isacharparenleft}{\kern0pt}{\isasymcdot}Bob{\isacharparenright}{\kern0pt}{\isacharparenright}{\kern0pt}{\isacartoucheclose}%
\isadelimproof
\ %
\endisadelimproof
\isatagproof
\isacommand{by}\isamarkupfalse%
\ eval%
\endisatagproof
{\isafoldproof}%
\isadelimproof
\endisadelimproof
}
\DefineSnippet{proposition:3aa6a594d1ecf19a-0}{%
\isacommand{proposition}\isamarkupfalse%
\ {\isacartoucheopen}{\isasymnot}prover\ {\isacharparenleft}{\kern0pt}riddle\ {\isasymrightarrow}\ {\isasymcdot}Bob{\isacharparenright}{\kern0pt}{\isacartoucheclose}%
\isadelimproof
\ %
\endisadelimproof
\isatagproof
\isacommand{by}\isamarkupfalse%
\ eval%
\endisatagproof
{\isafoldproof}%
\isadelimproof
\endisadelimproof
}
\DefineSnippet{end:7158f6523c28304a-0}{%
\isadelimtheory
\endisadelimtheory
\isatagtheory
\isacommand{end}\isamarkupfalse%
\endisatagtheory
{\isafoldtheory}%
\isadelimtheory
\endisadelimtheory
}

\newcommand{\Snippet}[1]{{%
  \newcount\i
  \i=0
  \loop
    \csname snippet--#1-\the\i\endcsname
    \advance \i 1
  \ifcsname snippet--#1-\the\i\endcsname
  \repeat
}}

\newcommand{\SnippetPart}[3]{{%
  \newcount\i
  \i=#1
  \loop
    \ifnum \i=#2
      \renewcommand{\isanewline}{}%
    \fi
    \csname snippet--#3-\the\i\endcsname
    \advance \i 1
    \ifnum \i>#2 {}
    \else \repeat
}}

\def\isacartoucheopen{\isatext{\raise.3ex\hbox{$\scriptscriptstyle\langle\,\,\,$}}}%
\def\isacartoucheclose{\isatext{\raise.3ex\hbox{$\scriptscriptstyle\,\,\,\rangle$}}}%

\begin{document}

\maketitle

\begin{abstract}
We present a novel approach for teaching logic and the metatheory of logic to students who have some experience with functional programming.
We define concepts in logic as a series of functional programs in the language of the proof assistant Isabelle/HOL.
This allows us to make notions which are often unclear in textbooks precise, to experiment with definitions by executing them, and to prove metatheoretical theorems in full detail.
We have surveyed student perceptions of our teaching approach to determine its usefulness and found that students felt that our formalizations helped them understand concepts in logic, and that they experimented with them as a learning tool.
However, the approach was not enough to make students feel confident in their abilities to design and implement their own formal systems.
Further studies are needed to confirm and generalize the results of our survey, but our initial results seem promising.
\end{abstract}

\section{Introduction}%
\label{sec:introduction}
Logic is the foundation on which all of mathematics and computer science rests, and many undergraduate computer science programs therefore include an introductory course on logic.
The logical systems introduced in such a course can be applied to databases, domain-specific languages, artificial intelligence, computer security, formal verification, and many other topics.
At the Technical University of Denmark (DTU) we teach logic alongside logic programming in Prolog in a late-stage undergraduate course in addition to discrete mathematics taught in the first semester of the study program.
Undergraduate courses like ours give students a basic understanding of logic and just enough knowledge to start applying basic logical systems in their work.
But for students who really need to work with logic and design their own systems, this is not enough: they also need a good understanding of the metatheory of logic, i.e.\ why the systems work and how to prove that they do.
Proofs about logical systems are fraught with possibilities for subtle mistakes of understanding, and it is our experience that many students never really ``get'' how the logical systems, and the proofs about them, work.
Additionally, many textbooks define logical systems in terms of informal set theory and omit parts of their proofs (sometimes sweeping major complexities such as binders and substitution under the rug), which leads to difficulties in actually applying the theory when implementing real systems.
For many students, implementing logical systems in their projects (and making sure that they are correct) can thus seem like an insurmountable challenge.

We have recently begun giving a graduate course (course number 02256) on automated reasoning with course material implemented in the proof assistant Isabelle/HOL.
This year the course started with 80 students.
The official course listing is available at \url{https://kurser.dtu.dk/course/02256}.

Isabelle/HOL is the higher-order logic version of the generic proof assistant Isabelle.
Briefly, we can consider higher-order logic as the sum of functional programming and logic.
Isabelle/HOL allows us to write definitions as functional programs and to formally prove properties of these programs.
The proof assistant continuously checks that our proofs are correct, thus making it impossible to neglect any complexities.
Additionally, Isabelle/HOL allows us to automatically export appropriate definitions into ``real'' functional languages such as Haskell, OCaml, Scala and Standard ML, whereby they can be integrated into larger systems.

By defining the logical systems we teach within Isabelle/HOL, we are thus able to give precise and executable definitions of every aspect of the systems, and our proofs of metatheoretical properties such as soundness and completeness of systems relate directly to these precise definitions and are verified by the proof assistant.
Our experience is that students appreciate this: if they are in doubt about what something means, they can look up a precise definition and even execute it on examples of their own choosing to gain further understanding, and this should also make it clear how to implement the concepts in practice.
To determine whether students actually find our approach useful, we survey student perceptions of the components of our course and of their own abilities and behaviour.
Our hypotheses are that:
\begin{enumerate}
    \item Concrete implementations in a programming language aid understanding of concepts in logic.
    \item Students experiment with definitions to gain understanding.
    \item Our formalizations make it clear to students how to implement the concepts in practice.
    \item Our course makes students able to design and implement their own logical systems.
    \item Prior experience with functional programming is useful for our course.
    \item Our course helps students gain proficiency in functional programming.
\end{enumerate}
In summary, we contribute:
\begin{itemize}
    \item functional implementations of several logical systems which can be used to teach topics such as sequent calculus, natural deduction, de Bruijn indices, and algorithms for automatic theorem proving.
    \item formally verified proofs of common properties such as soundness and completeness for the systems mentioned above.
    \item an evaluation of the usefulness of our approach based on surveying student perceptions.
\end{itemize}
In the next section, we survey related work.
In \cref{sec:approach} we explain our teaching approach using a small verified automated theorem prover as an example.
In \cref{sec:survey} we survey student perceptions of our teaching approach to evaluate its usefulness, before concluding in \cref{sec:conclusion}.

\section{Related work}

There are of course numerous textbooks on logic, and several which include implementations of many of their definitions.
Examples include books by Harrison~\cite{Harrison2009}, Ben-Ari~\cite{BenAri2001}, and Doets and van Eijck~\cite{Doets2012}, which all contain implementations in various programming languages.
Unfortunately, these implementations are not connected directly with the definitions and proofs in the books, and it is thus sometimes unclear how the programs really relate to the definitions in the text.
Additionally, the proofs in these books are not verified by a proof assistant, and are about the informal definitions, not the programs themselves.
Language, Logic and Proof~\cite{Plummer2011} is a textbook and accompanying software package containing a number of small proof assistants designed to aid in teaching logic.
While this means that students can rest assured that their exercise proofs are formally verified, the proofs in the book itself are not, and it is again sometimes unclear exactly how the software packages relate to the notions introduced in the text.
In our approach, every logical system and notion is formally defined, and every theorem is proven in Isabelle/HOL, which allows students to see exactly how to implement concepts and inspect every detail of the proofs.

Some textbooks on logic do have formally verified proofs of their theorems, but these are typically not introductions to logic, but instead introductions to proof assistants that just happen to introduce some particular logic they are working in.
One example is Coq'Art~\cite{CoqArt}, which is an introduction to the Coq proof assistant and contains many formally verified proofs about logic and other topics.
Unfortunately, such books, being written for a much more advanced audience, are not by themselves good introductions to logic, since they presuppose knowledge beyond the usual computer science program.
In our approach, on the other hand, we teach an introduction to logic without presupposing much beyond basic knowledge of functional programming.

The approach of formalizing definitions as functional programs has successfully been used for a number of other topics in computer science.
The Software Foundations series~\cite{Pierce:SFold} covers some basic logic, programming languages, formal verification, functional algorithms and separation logic with definitions and proofs in the Coq proof assistant.
Concrete Semantics~\cite{ConcreteSemantics} is a textbook on programming language semantics with definitions and proofs in Isabelle/HOL.
Functional Algorithms, Verified!~\cite{FAV2021} is an introduction to functional data structures and algorithms with definitions and proofs in Isabelle/HOL.
Verified Functional Programming in Agda~\cite{Stump2016} is an introduction to functional programming itself, as well as functional algorithms, using the Agda programming language, which supports proofs about the programs written in it.

We have previously written about the contents of our courses~\cite{FMTea}, about teaching with Isabelle/Pure and Isabelle/HOL~\cite{ThEdu19,ThEdu20,ThEdu21}, and about individual formalizations~\cite{NaDeA18,SPA18,CILC,Villadsen20}, but not about the specifics and benefits of teaching logic and metatheory using functional programming to aid understanding, and we have not previously surveyed student perceptions about our approach in a rigorous manner.

\section{A glimpse of the teaching approach}\label{sec:approach}
As mentioned, our approach is to define the logical systems which are our objects of study as functional programs in Isabelle/HOL.
Having done this, we can then define e.g.\ automatic theorem provers and other derived programs.
Isabelle/HOL then allows us to prove results about the implementations directly, e.g.\ that our automatic theorem provers are sound and complete.

\subsection{The structure of the course}

\begin{table}
    \centering
    \begin{tabular}{c|l}
        Weeks & Topics \\\hline\\[-1.5ex]
        1 -- 2  & Basic set theory, propositional logic, sequent calculus, automatic theorem proving \\[.5ex]
        3 -- 4  & Syntax and semantics of first-order logic, natural deduction, the LCF approach \\[.5ex]
        5 -- 6  & Isar, intuitionistic logic, foundational systems \\[.5ex]
        7 -- 8  & Proof by contradiction, classical logic, higher-order logic, type theory \\[.5ex]
        9 -- 10 & Proofs in sequent calculus \\[.5ex]
        11 -- 13 & Metatheory, prover algorithms, program verification
    \end{tabular}
    \caption{Week plan and topics of our course.}
    \label{tab:weekplan}
\end{table}

\autoref{tab:weekplan} contains a general overview of the topics we went through in our course in spring 2022.
Note that functional programming is integrated in more or less all topics.
For a more detailed explanation of the contents of the 2020 and 2021 versions of our course and its place in the overall curriculum, see \cite{FMTea}.
For a discussion of our approach to teaching formal methods with Isabelle/HOL, see \cite{IsabelleWorkshop2022}.

The course consists of lectures, exercise sessions, and assignments.
This year, the lectures were in-person, but some parts were recorded for later use by the students.
The exercise sessions were supervised by teaching assistants who were only available in-person.
The lectures primarily covered theoretical topics, but also included tutorials on Isabelle/HOL and related software.
The exercise sessions consisted primarily of programming and proving exercises in Isabelle/HOL, with occasional use of other software.
There were 6 assignments during the course, and students worked on them individually.
The assignments consisted of larger exercises as well as exercises similar to those in the exam.

This year, our graduate course on automated reasoning started with 80 students, but only 43 were left at the time of the survey.
This is not unexpected, since our university allows students to deregister for courses within the first month with no consequences.
This means that many students spend the first few weeks auditing many courses, and then choose which courses to stay registered to.
Almost all of the students registered at the time of the survey stayed on for the rest of the course and registered for the exam.
If a student registers for the exam then the course becomes binding for the student and the student must pass the course in order to graduate.
Three exam attempts are allowed.

The exam consisted of five problems each comprising two questions, and the students were given two hours to complete it.
One of the problems consisted of writing simple functions and proving properties about them in Isabelle/HOL.
The two questions were as follows:
\begin{itemize}
\item 
Using only the constructors $0$ and $\mathrm{Suc}$ and no arithmetical operators, define a recursive function $\mathrm{triple} :: \mathrm{nat} \Rightarrow \mathrm{nat}$ and prove $\mathrm{triple}~n = 3 * n$.
\item 
Using the usual operators $+$ and $-$, define two recursive functions $\mathrm{add42} :: \mathrm{int}~\mathrm{list} \Rightarrow \mathrm{int}~\mathrm{list}$ and $\mathrm{sub42} :: \mathrm{int}~\mathrm{list} \Rightarrow \mathrm{int}~\mathrm{list}$ that adds $42$ to each integer and subtracts $42$ from each integer, respectively, and prove $\mathrm{sub42}~(\mathrm{add42}~xs) = xs \land \mathrm{add42}~(\mathrm{sub42}~xs) = xs$. 
\end{itemize}
Both questions can be solved by defining the functions and instructing Isabelle to perform induction.
While these questions are not very difficult to solve, being able to prove properties of basic functional programs provides the basis for formalizing concepts in logic.
From these simple beginnings, we can build an understanding of more complicated functions and proofs such as the example we introduce in the next section.
Along the way we of course need to introduce both the logical concepts and the language and proof methods of the Isabelle proof assistant itself.

For a concrete example of the types of exam problems used, in particular on logic and the use of Isabelle, see the extended abstract \cite{ThEdu22}.

The course evaluations and the grade history can be found on the Information tab on the official course listing available at \url{https://kurser.dtu.dk/course/02256}.

\newcommand{\snip}[1]{\begin{footnotesize}\begin{isabelle}\Snippet{#1}\end{isabelle}\end{footnotesize}}

\begin{figure}

\snip{datatype:form}

\snip{primrec:semantics}

\snip{abbreviation:sc}

\snip{primrec:member}

\snip{lemma:member-iff}

\snip{primrec:common}

\snip{lemma:common-iff}

\snip{function:mp}

\snip{termination:mp}

\snip{theorem:main}

\snip{definition:prover}

\snip{corollary:13947c02cc06a4ca}

\caption{An example Isabelle/HOL development defining a propositional logic and an automatic theorem prover for it. We begin by defining formulas with propositions, falsity, and implication as a datatype, then define semantics of formulas as a function. We then define an automatic theorem prover (function \isa{mp}) for the system and prove that it terminates and is sound and complete. Note how every definition is a simple functional program and that Isabelle/HOL allows us to prove properties of these programs directly. Our students study this example very early on in our graduate course.}
\label{fig:isabelle-example}
\end{figure}

\begin{figure}
\begin{verbatim}
{-# LANGUAGE EmptyDataDecls, RankNTypes, ScopedTypeVariables #-}

module Scratch(Form, prover) where {

import Prelude ((==), (/=), (<), (<=), (>=), (>), (+), (-), (*), (/), (**),
  (>>=), (>>), (=<<), (&&), (||), (^), (^^), (.), ($), ($!), (++), (!!), Eq,
  error, id, return, not, fst, snd, map, filter, concat, concatMap, reverse,
  zip, null, takeWhile, dropWhile, all, any, Integer, negate, abs, divMod,
  String, Bool(True, False), Maybe(Nothing, Just));
import qualified Prelude;

data Form a = Pro a | Falsity | Imp (Form a) (Form a);

member :: forall a. (Eq a) => a -> [a] -> Bool;
member uu [] = False;
member m (n : a) = m == n || member m a;

common :: forall a. (Eq a) => [a] -> [a] -> Bool;
common uu [] = False;
common a (m : b) = member m a || common a b;

mp :: forall a. (Eq a) => [a] -> [a] -> [Form a] -> [Form a] -> Bool;
mp a b (Pro n : c) [] = mp (n : a) b c [];
mp a b c (Pro n : d) = mp a (n : b) c d;
mp uu uv (Falsity : uw) [] = True;
mp a b c (Falsity : d) = mp a b c d;
mp a b (Imp p q : c) [] = mp a b c [p] && mp a b (q : c) [];
mp a b c (Imp p q : d) = mp a b (p : c) (q : d);
mp a b [] [] = common a b;

prover :: forall a. (Eq a) => Form a -> Bool;
prover p = mp [] [] [] [p];

}
\end{verbatim}
\caption{Haskell code generated by Isabelle/HOL from the example in Figure~\ref{fig:isabelle-example}. Note that it is essentially a direct translation of the relevant Isabelle/HOL definitions, and that the proofs are not included. While the code is not pretty, a module such as this one can easily be integrated with other (handwritten) code to create a full application. This provides an example of how to implement and use the logical systems we cover in our course in practice.}
\label{fig:haskell-example}
\end{figure}

\subsection{Logical systems and provers as functional programs}
As a simple example, \cref{fig:isabelle-example} contains a definition of classical propositional logic.
We define a datatype of formulas (this is the so-called deep embedding approach), then introduce semantics as a function from truth value assignments (interpretations) and formulas to truth values (we also define a semantics function for sequents, \isa{sc}).
We can then define an automatic theorem prover (function \isa{mp} for micro prover) for the system which works by breaking formulas up using a system similar to a sequent calculus.
All of these definitions are simple functional programs, and students should thus be able to understand them quite quickly when the ideas behind them are explained.

The programs can be executed directly inside of Isabelle/HOL, but can also be exported to standard functional programming languages, which allows us to use the concepts in practice.
The Haskell program in \cref{fig:haskell-example} has been automatically generated by Isabelle/HOL from the definitions in \cref{fig:isabelle-example}.

Note that only the functions themselves, and not the proofs about them, are present in the Haskell code.
The idea behind this approach is that the difficult parts of a program can be formally verified using Isabelle/HOL, then exported to a ``normal'' programming language as modules ready for integration into the overall program.
This is similar to the LCF approach~\cite{Harrison2014,PaulsonNW19} to theorem proving and the ``hexagonal architecture'' pattern common in object-oriented and functional programming~\cite{Hexagonal2005}, both of which advocate for programs consisting of a core application which interacts with users and other programs through an outer layer.
In this case, the core can be implemented and verified in Isabelle/HOL, then exported, while the communication layer can be implemented in the target programming language.

\subsection{Metatheory as properties of programs}
Once we have defined a logical system in Isabelle/HOL, we can begin proving results about it.
In \cref{fig:isabelle-example}, we first prove that the functions \isa{member} and \isa{common} are correct by simple structural induction.
Note that the proofs very much resemble the classic ``The proof is by induction'' often found in textbooks, but that the proof has been verified by Isabelle/HOL.
This allows the reader to rest easy knowing that there is no hidden complexity in the proof.

We then prove termination of our prover by noting that the sum of the sizes of the two last arguments decrease.
Isabelle/HOL requires that all functions are total, and in this case the proof is complicated enough that we have to prove it manually, but simple enough that we can do so easily.
Next, we prove that our automatic theorem prover returns true if and only if the provided sequent is valid (theorem \isa{main}).
This proof is by induction, and the automation in Isabelle/HOL is enough to handle all cases.
Finally, we conclude that our prover returns true for a formula exactly when it is valid (true in all interpretations), which means that the prover is sound and complete.

In this fashion, we can prove metatheoretical results about several logical systems used in our course, and thus showcase the proof techniques needed.
We are also able to explore and prove equivalences between different logical systems by proving that the programs implementing them return the same results.

\section{Student perceptions of the teaching approach}\label{sec:survey}
To attempt to shed some light on whether our approach is actually useful, we have conducted a survey study to analyze student perceptions about our course.
The survey consisted of a single self-administered questionnaire, which all students following the course were invited to fill in during the tenth week of the course.

\subsection{Methods}\label{sec:methods}
The study was designed to address the following six hypotheses:
\begin{enumerate}
    \item Concrete implementations in a programming language aid understanding of concepts in logic.
    \item Students experiment with definitions to gain understanding.
    \item Our formalizations make it clear to students how to implement the concepts in practice.
    \item Our course makes students able to design and implement their own logical systems.
    \item Prior experience with functional programming is useful for our course.
    \item Our course helps students gain proficiency in functional programming.
\end{enumerate}

To address these hypotheses, we developed 12 questions to measure student perception of the course and their own behaviour and abilities.
The questions can be seen on the left in \cref{fig:midterm-answers} or in the appendix.
The questions were developed following evidence-based best practices for questionnaire item question design as described in \cite{Artino2014}, but the consistency and reproducibility of answers to the questionnaire items have not been tested.

Questions were close-ended and answers were given in terms of perception on one of three Likert-type \cite{Likert1932} scales (multiple answers not possible) based on established best practice response options \cite{Artino2014}:
\begin{description}
\item[An importance scale] consisting of the levels: "Not important", "Slightly important", "Moderately important", "Quite important", and "Essential".
\item[A frequency scale] consisting of the levels: "Almost never", "Once in a while", "Sometimes", "Often", and "Almost always".
\item[A confidence scale] consisting of the levels: "Not at all confident", "Slightly confident", "Moderately confident", "Quite confident", and "Completely confident".
\end{description}
Since the questionnaire consisted almost exclusively of attitudinal questions, we did not include a "Don't know" option for any question since this incentivizes partial responses \cite{Harlacher2016}.
``Don't know'' responses were thus not possible, and students were required to fill in the entire questionnaire to submit it, which means that partial responses were not possible.

Since the survey consisted of a single questionnaire which was administered only once, we have only one sample (but note that question 10 asks about previous perception, which we analyze as a separate sample).
The survey population was all 43 students following the course during course week 10.
To preserve student anonymity, the sampling frame does not include auxiliary information such as gender, age, grades, etc.
Specifically, this means that we cannot connect student answers to the questionnaire with measures of learning outcomes.
We expect that there is some self-selection bias, since some students officially follow the course but do not actually show up to class, and are thus unlikely to answer the questionnaire.
Additionally, we expect that students who either really like or really dislike the course will be more likely to answer the questionnaire, which could also result in some self-selection bias.

The questionnaire was administered using an online surveying solution provided by our university.
This let us ensure that each student was only able to fill in the questionnaire once, and that participating students were anonymous.
Participation in the survey was optional and not compensated financially or otherwise.
The students were asked to participate in several lectures and through email.

\subsection{Data analysis}
The questionnaire has been designed to answer the hypotheses listed above.
In this section, we will describe how we intend to confirm or reject each of the hypotheses based on the answers to the questions.
Since we did not allow partial responses, we do not need to handle missing data for individual questions.

All of our data analysis was performed using the R environment for statistical computing \cite{R2022}.
The data from the questionnaire was pre-processed using the readr package \cite{readr2022}.
The analysis script is available at \url{https://github.com/fkj/tfpie-2022-statistics}.

\subsubsection{Concrete implementations in a programming language aid understanding of concepts in logic}
Questions 1--3 ask about the perceived importance of the three elements of the course.
We would like to compare the relative importance of the three elements to determine whether the implementations play an important part in gaining understanding.

If we had a measure of student learning outcomes, it would be interesting to perform a relative importance analysis on the three factors, but due to the anonymity of the questionnaire, we are not able to couple it to e.g.\ grades.
Instead, we simply qualitatively compare the results of the questionnaire.

\subsubsection{Students experiment with definitions to gain understanding}
Question 4 asks whether students think it is important to experiment with their own examples when learning about a new concept.
Question 5 asks whether students actually evaluate concrete examples to gain understanding when they are using Isabelle.
We can look at the frequencies and median answers to gain an understanding of student behaviour.

\subsubsection{Our formalizations make it clear to students how to implement the concepts in practice}
Question 6 asks directly about how the students perceive their abilities to do this.
We can look at the frequencies and median answer to gain an understanding of student perceptions.

\subsubsection{Our course makes students able to design and implement their own logical systems}
The meaning of this hypothesis is a bit vague, so we divide it into multiple concrete questions.
Question 7 asks about perception of ability to design a formal system to solve a practical problem.
Question 8 asks about perception of ability to implement a formal system in any programming language (without proofs).
Question 9 asks about perception of ability to prove a formal system correct.
We can look at the frequencies and median answers to gain an understanding of student perceptions.

\subsubsection{Prior experience with functional programming is useful for our course}
Question 10 asks about perceived ability with functional programming before starting the course and we will measure the association with perceived ability during the exercises
and assignments in the course, which we ask about in question 11.
Since both categories are ranked from low to high, there is no possibility of multiple choices, and we are interested in question 10 as a predictor for question 11, we will use Somers' Delta statistic \cite{DescTools2021} to measure the association \cite{Mangiafico2016}.
We can also look at the frequencies and median answers to gain an understanding of student perceptions.

\subsubsection{Our course helps students gain proficiency in functional programming}
Question 10 asks about perceived ability before starting the course, while question 12 asks about perceived ability towards the end of the course.
We can test whether the probability of high perceived ability has increased during the course by comparing the two answers for each student.
Since we have two groups with paired observations (i.e.\ the same students twice), we can perform a two-sample paired signed-rank test \cite{Mangiafico2016} (i.e.\ a paired Wilcoxon signed-rank test) to determine whether there is a significant difference in perceived functional programming ability before and after the course \cite{Mangiafico2016}.
The effect size of the paired Wilcoxon signed-rank test can be measured by the matched-pairs rank biserial correlation coefficient \cite{Mangiafico2016}.

\subsection{Results}
21 students completed the entire questionnaire (48.84\% response rate).
The margin of error is thus below 15.48\% (95\% CI, $p$ set to 0.5 to obtain worst-case margin).
A summary of the answers can be seen in \cref{fig:midterm-answers}, and the full data set is available in the appendix.
Based on these answers, we will present the result of the analyses mentioned for each hypothesis above.

\begin{figure}
    \centering
    \includegraphics[width=\textwidth]{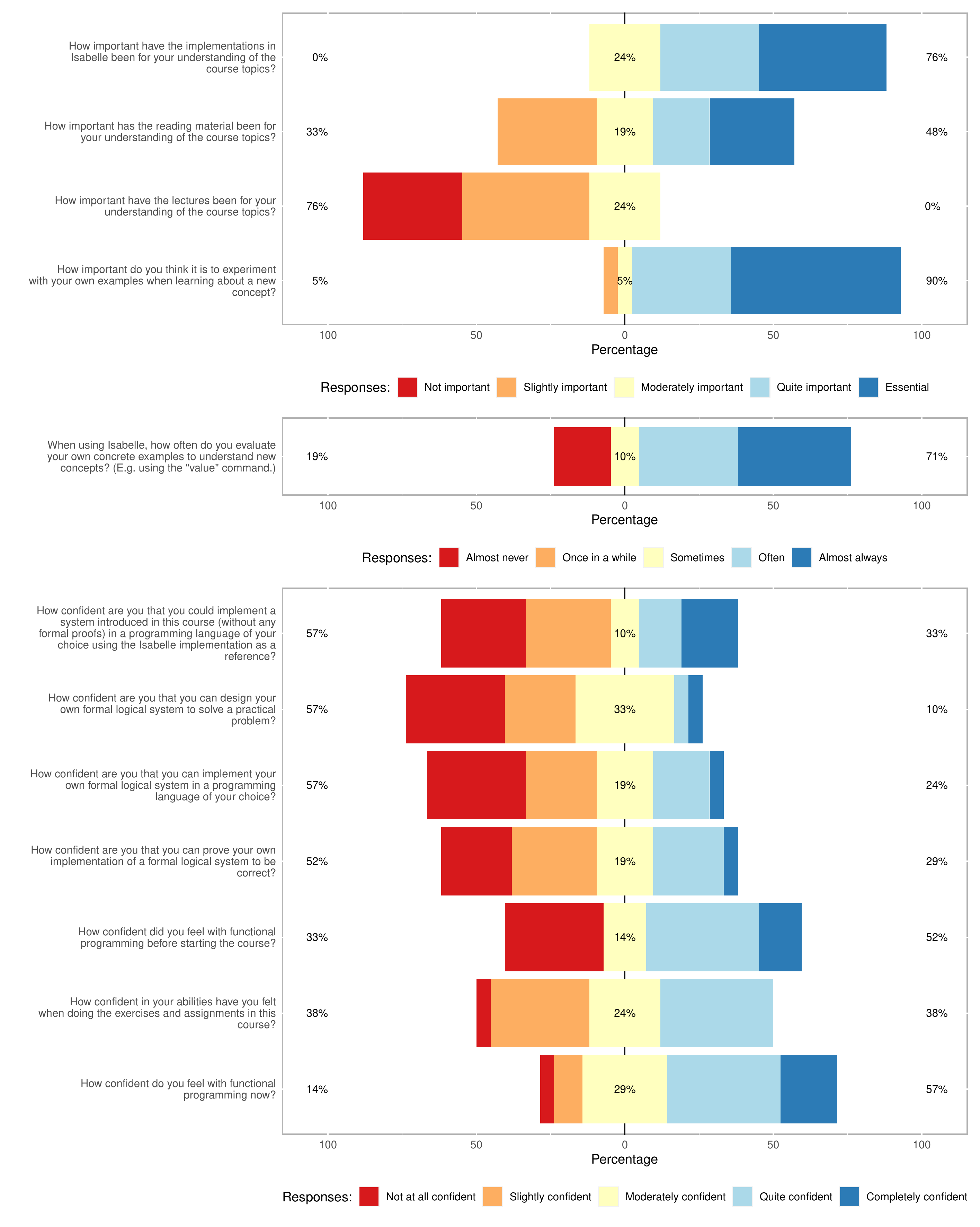}
    \caption{Summary of answers to the student self-evaluation questionnaire (n = 21). See the appendix for the full data set.}
    \label{fig:midterm-answers}
\end{figure}

\subsubsection{Concrete implementations in a programming language aid understanding of concepts in logic}
Comparing the answers to question 1--3, we see that the median student thinks that:
\begin{enumerate}
    \item the implementations in Isabelle are ``Quite important'' for their understanding of the course topics
    \item the reading material is ``Moderately important'' for their understanding of the course topics
    \item the lectures are ``Slightly important'' for their understanding of the course topics
\end{enumerate}
This indicates that students believe that the implementations in Isabelle are the most important component of the course when it comes to their understanding of the course topics.
While we do not have access to an association between these beliefs and actual learning outcomes, this data suggests that the hypothesis is plausible.

\subsubsection{Students experiment with definitions to gain understanding} \label{sec:results:hyp2}
Looking at the answers to question 4, we see that all students find experiments with their own examples at least slightly important, while most students find them at least quite important.
The median student finds experiments with their own examples ``Essential'' when learning about a new concept.

Looking at the answers to question 5, we see that there seems to be two distinct groups: one group which ``Almost never'' evaluates concrete examples in Isabelle, and a group which does so quite often.
The median student evaluates concrete examples in Isabelle ``Quite often'', which confirms our hypothesis.

\subsubsection{Our formalizations make it clear to students how to implement the concepts in practice} \label{sec:results:hyp3}
Looking at the answers to question 6, we see that most students are not at all, or only slightly confident, but that there are also a number of students who are quite, or completely confident.
The median student, however, is only ``Slightly confident'' that they could implement the concepts in practice.
This suggests to reject the hypothesis.

\subsubsection{Our course makes students able to design and implement their own logical systems} \label{sec:results:hyp4}
Looking at the answers to question 7, we see that the general trend is that student do not feel that they have the ability to design formal systems to prove practical problems.
The median student is only ``Slightly confident'' in their ability to do this.
This suggests to reject the hypothesis.

Looking at the answers to question 8, we see that the general trend is that students do not feel that they have the ability to implement their own logical systems (in any programming language).
The median student is only ``Slightly confident'' in their ability to do this.
This suggests to reject the hypothesis.

Looking at the answers to question 9, we see that the general trend is that students are only slightly to moderately confident that they have the ability to prove a formal system correct.
The median student is only ``Slightly confident'' in their ability to do this.
This suggests to reject the hypothesis.

Overall, the data suggests to reject the hypothesis.

\subsubsection{Prior experience with functional programming is useful for our course} \label{sec:results:hyp5}
Looking at the answers to question 10, we see that there is an approximately normal distribution of students who felt moderately to completely confident as well as a second mode of students who were not at all confident with functional programming before starting the course.
We know that this second mode consists of students who did not have any experience with functional programming prior to the course (although it was an explicit prerequisite).
The median student felt ``Quite confident'' with functional programming before starting the course.

Looking at the answers to question 11, we see that nobody felt completely confident, while most students felt slightly to quite confident when doing the exercises and assignments in the course.
The median student felt ``Moderately confident'' when doing the exercises and assignments.

Somers' Delta statistic for the association of question 10 as a predictor for question 11 is 0.3642 (95\% CI $[0.0038, 0.7247]$), which suggests a small to moderate association between students who felt confident with functional programming before starting the course and students who felt confident during the exercises and assignments \cite{Ferguson2009}.
This suggests to confirm the hypothesis.

\subsubsection{Our course helps students gain proficiency in functional programming} \label{sec:results:hyp6}
Looking at the answers to question 10, we see that a number of students had no confidence with functional programming before starting the course (presumably due to them missing the functional programming prerequisite).
Looking at the answers to question 12, we see that after the course, we obtain what looks more like a normal distribution centered around the "Quite confident" answer (although it is quite skewed).

The two-sample paired signed-rank test gives V = 2 (95\% CI $[-3.5, 0.00]$), with $p = 0.0498$.
Since $p \leq 0.05$, we conclude that there is a statistically significant difference in perceived functional programming ability before and after the course \cite{Mangiafico2016}.
The effect size as measured by the matched-pairs rank biserial correlation is -0.857 (95\% CI $[-1.00, -0.333]$), which suggests that the course has a large positive effect on perceived functional programming ability \cite{Mangiafico2016}.
This suggests to confirm the hypothesis.

\subsection{Discussion}
We begin with a summary of our results:
\begin{description}
    \item[Plausible:] Concrete implementations in a programming language aid understanding of concepts in logic.
    \item[Confirmed:] Students experiment with definitions to gain understanding.
    \item[Rejected:] Our formalizations make it clear to students how to implement the concepts in practice.
    \item[Rejected:] Our course makes students able to design and implement their own logical systems.
    \item[Confirmed:] Prior experience with functional programming is useful for our course.
    \item[Confirmed:] Our course helps students gain proficiency in functional programming.
\end{description}

Our first result suggests that our students feel that our teaching approach does help them understand concepts in logic.
Unfortunately, our study does not have the data to confirm that they actually do, since we do not have access to assessments of learning outcomes linked to student responses.
We also confirm that students do in fact use the formal definitions as learning tools by experimenting with examples and definitions within Isabelle.

On the other hand, students did not feel that our formalizations were enough to make it clear how to design and implement their own logical systems, or the concepts in the course in general.
While we did not expect that students would feel very confident doing this (since formal verification is generally very difficult), we had still hoped for more confident answers.
This suggests that we should attempt to find ways to improve our course in this respect.
We also note that students do not seem to think that our lectures are very useful, which suggests that we should consider whether they are necessary or if even more time should be spent on exercises.
By reducing the time spent on lectures, it may also be possible to obtain time for some project work, which may help students become more confident in designing and implementing their own logical systems.

We also found that students with more functional programming experience before starting the course felt more confident during the course, which is as expected.
Our data also suggests that our course has a large positive effect on self-perceived functional programming confidence (though this may be a result of some students having no prior functional programming experience, see \cref{sec:post-hoc-analysis}).

Our study is quite limited in scope, and the confidence intervals on our results are thus not very narrow.
A larger study is needed to obtain precise knowledge of effect sizes.
Since students participated in the survey of their own volition and with no compensation, we expect that our results also suffer from some self-selection bias.
Further studies are needed to determine whether our results generalize.

\subsubsection{Post hoc exploratory analyses}\label{sec:post-hoc-analysis}
Having seen the answers to our survey, we notice some phenomena that could be interesting to explore further.
Note that the analyses in this section are post hoc, i.e.\ they have been formulated after seeing the data, and that further studies are thus necessary to confirm any effects described in this section.

\paragraph{In \cref{sec:results:hyp2},} we noted that the answers to question 5 seem to contain two distinct groups.
We can think of two possible reasons that a group of students almost never experiment with concrete examples in Isabelle: either there is a group of students who simply had not understood that this is possible, or some students do not find experimentation valuable enough to actually do it outside of the abstract.
To investigate this further, we can look at the distribution of answers to whether students think that experimentation is important for students who answered that they almost never evaluate concrete examples in Isabelle and students who answered otherwise.
We can split the data set in "Almost never" and "Other" and look at the distribution of how important students think experimentation is for each category.
By doing this, we see that the shapes of the two distributions do not seem to differ significantly.

We can also measure the association between students who think that experimentation is important and students who often evaluate concrete examples in Isabelle.
Since both categories are ranked from low to high, there is no possibility of multiple choices, and we are interested in question 4 as a predictor for question 5, we will use Somers' Delta statistic \cite{DescTools2021} to measure the association \cite{Mangiafico2016}.
The statistic is $-0.1039$ (95\% CI $[-0.4777, 0.2699]$), which shows a small negative association between the variables, i.e.\ that students who find experimentation more important evaluate concrete examples slightly less often than those who find it less important.
This is surprising.
If we assume that the answers of "almost never" are due to students not knowing about the feature, and thus remove them from the data set as special cases, the association becomes $-0.0814$ (95\% CI $[-0.5046, 0.3418]$), which again shows a small negative association.
While there is not enough data to give good confidence intervals, it seems like there is no statistically significant association between the perceived importance of experimentation and the frequency of evaluating concrete examples in Isabelle.
Note, however, that students still generally tend to believe that experimentation is important, and that they generally tend to evaluate their own concrete examples in Isabelle.
One possible explanation is that some students who believe that experiments are essential are used to doing experiments with pen and paper, and thus do this instead of using Isabelle to evaluate examples, but we do not have any data to investigate this hypothesis in this study.

\paragraph{In \cref{sec:results:hyp3},} we noted that the answers to question 6 seem to contain two distinct groups.
The bimodality is interesting, since we know that many students did not have experience with functional programming prior to our course (even though it was an explicitly stated prerequisite).
We can test whether there is an association between perceived functional programming ability and perceived ability to implement the concepts in practice.
We have data for perceived programming ability both before and after the course, so we can analyze both.
Since all categories are ranked from low to high, there is no possibility of multiple choices, and we are interested in questions 10 and 12 as predictors for question 6, we will use Somers' Delta statistic \cite{DescTools2021} to measure the associations \cite{Mangiafico2016}.
For previous functional programming ability, the statistic is 0.194 (95\% CI $[-0.139, 0.527]$).
This suggests that there is an association, but that it is barely significant \cite{Ferguson2009}.
For current functional programming ability, the statistic is 0.394 (95\% CI $[0.0807, 0.707]$).
This suggests a small to moderate, but statistically significant association \cite{Ferguson2009}.

We conclude that it seems that our formalizations may help students understand how to implement the concepts in practice, but only significantly so if they were confident functional programmers after following the course.
Further studies are needed to confirm this effect, since the analysis above is post hoc.

\paragraph{In \cref{sec:results:hyp4},} we concluded that students do not generally feel confident in their abilities to implement their own logical systems.
It could be interesting to measure the association between perceived functional programming ability and perceived ability to implement logical systems.
We have data for perceived programming ability both before and after the course, so we can analyze both.
Since both categories are ranked from low to high, there is no possibility of multiple choices, and we are interested in questions 10 and 12 as predictors for question 8, we will use Somers' Delta statistic \cite{DescTools2021} to measure the associations \cite{Mangiafico2016}.
For previous functional programming ability, the statistic is 0.108 (95\% CI $[-0.259, 0.474]$).
This suggests that there is no statistically significant association \cite{Ferguson2009}.
For current functional programming ability, the statistic is 0.419 (95\% CI $[0.0445, 0.794]$).
This suggests a small to moderate association \cite{Ferguson2009}.

We conclude that it seems that our formalizations may help students understand how to implement logical systems, but only if they are confident functional programmers after the course, and only slightly so.

\paragraph{In \cref{sec:results:hyp5},} we noted that the answers to question 10 seem to have two modes.
If we look at the two modes independently, the students with the functional programming prerequisite tend to feel moderately to quite confident while most students without the functional programming prerequisite felt only slightly confident.
This also supports the original hypothesis.

We can try to remove the students with no functional programming experience and measure the association between perceived ability with functional programming before starting the course and perceived ability during the exercises and assignments in the course again to see whether the course requires advanced functional programming skills.
We will again use Somers' Delta statistic to measure association \cite{Mangiafico2016}.
The value is 0.0000 (95\% CI $[-0.5019, 0.5019]$), which we interpret as meaning that only basic knowledge of functional programming seems to be needed in our course.
This tracks well with our design choices in our systems, where we deliberately avoid ``advanced'' concepts such as folds in our implementations and exercises in an effort to obtain this result.

\paragraph{In \cref{sec:results:hyp6},} we found that the course has a significant positive effect on perceived functional programming ability.
We can try to remove the students with no functional programming experience to see whether there is still a significant difference for those students who already have functional programming experience.
We again perform a two-sample paired signed-rank test (i.e. a paired Wilcoxon signed-rank test) to determine whether there is a significant difference in perceived functional programming ability before and after the course \cite{Mangiafico2016}.
The test shows V = 1, with $p = 1$.
This means there is no statistically significant difference in perceived functional programming ability for the group of students who already have functional programming experience.
In fact, by inspection of the data set we see that every student has answered exactly the same for both questions, except one student who moved from feeling "Quite confident" to feeling only "Moderately confident" after having taken the course.
The lack of effect on perceived functional programming ability for students who already have some experience is to be expected since our course does not utilize any advanced concepts in functional programming.

\subsection{Future work}
Our results suggest that our approach seems to aid student understanding of logic, but that students do not feel that our formalizations are enough to make it clear how to implement the concepts in practice, and that students do not feel capable of designing and implementing their own logical systems after the course.
This suggests that we should attempt to find ways to make it more clear how to do this in future iterations of the course.
One possible way to do this would be to introduce a project about implementing some system to solve a practical problem, but this would almost certainly require us to extend the course load.
It may also be possible to replace some assignments and exercises during the course with more practically oriented ones.
Additionally, we may be able to reduce the time spent on lectures, since students do not seem to find them very important for their learning outcomes.

Our survey consisted almost entirely of attitudinal questions, which means that we are forced to believe student self-perceptions to some degree.
We could improve this aspect by assessing e.g.\ understanding of logic and experience with functional programming by learning outcome measures such as previous grades or tests during the survey.
Adding tests of aptitude to the survey would make completing it a much larger endeavour, however, which might decrease the number of students willing to participate and introduce additional self-selection bias.
Including measures such as previous grades would require us to handle non-anonymous data, which makes the survey a much larger undertaking, since we would then need to collect informed consent letters, obtain institutional review board approval, etc.
The same is true for demographic data such as gender, age, etc.

The generalizability of our survey may be doubted, since it may have issues of self-selection bias, and since our sample size was not large enough to obtain a narrow margin of error.
Self-selection bias may be decreased by screening participants to ensure that our sample is representative of the population, but this will also require collection of demographic data and possibly previous grades, and will most likely reduce the sample size.
Another way to reduce self-selection bias is to collect data on every student, but this has obvious ethical issues.
In the future, our survey could be repeated on other cohorts to increase the generalizability of our results.

Since our course is new, we have no data from previous pen-and-paper based courses at our university to compare our approach to.
Performing such a comparison in a rigorous way would also be difficult, since it is not obvious how to choose a test of student learning outcomes that is comparable for both pen-and-paper approaches and our approach.

Finally, it may be interesting to conduct further studies to determine whether the effects discussed in \cref{sec:post-hoc-analysis} actually exist.
We have noted the following interesting questions which arise from phenomena in our sample, but which, being the result of post hoc analyses, cannot be confirmed from the answers to the present survey:
\begin{itemize}
    \item Why do students who think experimentation is important do it less? Do they do it on paper instead?
    \item Does functional programming experience play a significant role in understanding of how to implement concepts in practice?
    \item Does functional programming experience play a significant role in understanding of how to design and implement one's own logical systems?
    \item Does our course have a positive effect on functional programming skill for students who are already confident functional programmers?
\end{itemize}

\section{Conclusion}\label{sec:conclusion}
We have presented our approach for teaching logic and metatheory using functional programming and demonstrated how logical concepts can be defined as simple functional programs about which we can prove metatheoretical results in a natural way.
We have surveyed our students to determine their perceptions of our teaching approach.
Our survey shows that students feel that our formalizations do aid them in understanding concepts in logic.
We also confirm that students do in fact use the formal definitions as learning tools by experimenting with examples and definitions within Isabelle.
On the other hand, students did not feel that our formalizations were enough to make it clear how to design and implement their own logical systems, or the concepts in the course in general.
This suggests that we should attempt to find ways to make this clearer in future iterations of the course.
Our survey also revealed a number of interesting phenomena which may be interesting to pursue further in future studies.
Future studies are also needed to improve the generalizability of our results.
We believe that our study design can serve as a useful methodology for such further studies.

For now, our course material consists of a series of lecture notes and corresponding formalizations and student exercise templates in Isabelle/HOL, and we combine this with excerpts from textbooks and tutorials.
It would be great to have a comprehensive and coherent textbook written with this style of teaching in mind, but our course material has not yet stabilized enough for us to consider writing one.
We would like to make clear that a book containing nothing but programs is not what we advocate; instead, we would like a textbook containing explanations based on, and corresponding to, the precise definitions in the functional programs, such that readers can always clarify any doubts about the ``intuitive'' explanations given in the text by referring to the programs implementing the definitions.
Similarly, we do not believe that textbooks should contain meticulous proofs going into every detail of every case, but that such proofs should be available to the reader if they are not convinced by the abbreviated arguments given in the text.

\section*{Acknowledgements}
We thank Agnes Moesgård Eschen, Asta Halkjær From, Alceste Scalas, Michael Reichhardt Hansen and Simon Tobias Lund for comments on drafts.
We also thank the anonymous reviewers for their useful suggestions.

\bibliographystyle{eptcs}
\bibliography{references}

\section*{Appendix: Survey data}
To save space, the survey data has been compressed using numeral encodings.
The questions are numerically encoded according to the following table, which also shows the Likert-type scale used for their response options (see \autoref{sec:methods} for details).
\bigskip
\begin{center}
\begin{tabular}{@{}c|p{13cm}|l@{}}
\# & Question & Scale \\\hline
1 & How important have the implementations in Isabelle been for your understanding of the course topics?     & Importance \\
2 & How important has the reading material been for your understanding of the course topics?                 & Importance \\
3 & How important have the lectures been for your understanding of the course topics?                        & Importance \\
4 & How important do you think it is to experiment with your own examples when learning about a new concept? & Importance \\
5 & When using Isabelle, how often do you evaluate your own concrete examples to understand new concepts? (E.g. using the "value" command.)                                                                  & Frequency \\
6 & How confident are you that you could implement a system introduced in this course (without any formal proofs) in a programming language of your choice using the Isabelle implementation as a reference? & Confidence \\
7 & How confident are you that you can design your own formal logical system to solve a practical problem?                                                                                                   & Confidence \\
8 & How confident are you that you can implement your own formal logical system in a programming language of your choice?                                                                                    & Confidence \\
9 & How confident are you that you can prove your own implementation of a formal logical system to be correct? & Confidence \\
10 & How confident did you feel with functional programming before starting the course?                         & Confidence \\
11 & How confident in your abilities have you felt when doing the exercises and assignments in this course?     & Confidence \\
12 & How confident do you feel with functional programming now?                                                 & Confidence
\end{tabular}
\end{center}
\bigskip
\noindent
The responses to our questionnaire are numerically encoded in the following table according to the position on the Likert scale of the question, with 1 being the lowest possible level and 5 being the highest possible level.
\bigskip
\begin{center}
\begin{tabular}{@{}c|l|l|l|l|l|l|l|l|l|l|l|l|l|l|l|l|l|l|l|l|l@{}}
Question \# & \multicolumn{21}{c}{Answers (each column contains the answers from a single student)} \\\hline
1 & 5 & 4 & 4 & 5 & 3 & 5 & 3 & 4 & 5 & 5 & 4 & 4 & 3 & 4 & 4 & 5 & 5 & 5 & 3 & 5 & 3 \\
2 & 5 & 5 & 3 & 3 & 4 & 5 & 4 & 3 & 5 & 2 & 4 & 2 & 5 & 2 & 2 & 2 & 3 & 2 & 5 & 2 & 4 \\
3 & 3 & 2 & 2 & 3 & 2 & 2 & 2 & 3 & 2 & 2 & 1 & 1 & 3 & 1 & 1 & 2 & 1 & 1 & 1 & 3 & 2 \\
4 & 5 & 2 & 4 & 4 & 5 & 5 & 4 & 5 & 4 & 5 & 5 & 5 & 4 & 4 & 4 & 3 & 5 & 5 & 5 & 5 & 5 \\
5 & 5 & 5 & 4 & 5 & 4 & 4 & 4 & 1 & 4 & 5 & 3 & 1 & 5 & 3 & 5 & 1 & 4 & 1 & 5 & 5 & 4 \\
6 & 5 & 2 & 1 & 1 & 5 & 3 & 1 & 5 & 4 & 1 & 2 & 2 & 2 & 2 & 4 & 1 & 1 & 2 & 3 & 4 & 5 \\
7 & 5 & 1 & 1 & 1 & 3 & 3 & 1 & 3 & 4 & 2 & 2 & 2 & 3 & 2 & 1 & 3 & 1 & 1 & 3 & 3 & 2 \\
8 & 5 & 1 & 1 & 1 & 3 & 3 & 1 & 4 & 4 & 2 & 2 & 1 & 2 & 2 & 1 & 4 & 1 & 2 & 3 & 4 & 3 \\
9 & 5 & 1 & 1 & 1 & 2 & 4 & 2 & 4 & 3 & 2 & 3 & 2 & 3 & 4 & 4 & 3 & 1 & 1 & 4 & 2 & 2 \\
10 & 1 & 3 & 4 & 3 & 5 & 4 & 1 & 4 & 5 & 1 & 4 & 5 & 4 & 4 & 3 & 4 & 1 & 1 & 1 & 1 & 4 \\
11 & 3 & 4 & 4 & 3 & 3 & 4 & 1 & 4 & 4 & 2 & 4 & 2 & 3 & 3 & 2 & 2 & 2 & 4 & 2 & 2 & 4 \\
12 & 5 & 3 & 4 & 3 & 5 & 4 & 2 & 4 & 5 & 3 & 3 & 5 & 4 & 4 & 3 & 4 & 1 & 3 & 4 & 2 & 4
\end{tabular}
\end{center}

\end{document}